\documentclass[a4paper,amsmath,amssymb,color,nofootinbib,preprint,tightenlines]{revtex4-2}
\usepackage{graphicx}
\usepackage[dvipsnames]{xcolor}
\usepackage[mathscr]{euscript}
\usepackage{slashed}
\usepackage{tablefootnote}
\usepackage[normalem]{ulem}
\usepackage{cancel}
\usepackage{subcaption}
\usepackage{braket} 
\usepackage[colorlinks=true,pdfstartview=FitV,
citecolor=blue,linkcolor=purple,urlcolor=Sepia]{hyperref}
\usepackage[capitalise]{cleveref}
\usepackage{orcidlink}
\begin{document}
	\baselineskip=17pt \parskip=3pt
	
	\newcommand{\XGH}[1]{{\color{red}XGH: #1}}
	\newcommand{\XDM}[1]{{\color{orange}XDM: #1}}
	\newcommand{\GV}[1]{{\color{blue}GV: #1}}
	\newcommand{\JT}[1]{{\color{purple}JT: #1}}
	\begin{center}
		\textbf{{\large A light DM model for large $B \to K + \mbox{invisible}$ and $K \to \pi + \mbox{invisible}$ decays and its implications for $B_s-\bar B_s$ mixing and neutron EDM}} 
	\end{center}
	
	\author{Xuan Hong$,^{1,2,}$\footnote{\email{hxxh123@sjtu.edu.cn}} 
		Xiao-Gang He\,\orcidlink{0000-0001-7059-6311}$,^{1,2,}$\footnote{\email{hexg@sjtu.edu.cn}} 
		and Ming-Wei Li\,\orcidlink{0000-0003-2522-9795}$,^{1,2,}$\footnote{\email{limw2021@sjtu.edu.cn}
		} 
		\vspace{1ex} \\ \it
		$^1$State Key Laboratory of Dark Matter Physics, Tsung-Dao Lee Institute \& School of Physics and Astronomy,
		Shanghai Jiao Tong University, Shanghai 201210, China \vspace{1ex} \\
		$^2$Key Laboratory for Particle Astrophysics and Cosmology (MOE) \& Shanghai Key Laboratory for Particle Physics and Cosmology,
		Tsung-Dao Lee Institute \& School of Physics and Astronomy, Shanghai Jiao Tong University, Shanghai 201210, China \vspace{1ex} \\
		\rm Abstract 
		\vspace{8pt} \\
		\begin{minipage}{0.99\textwidth} \baselineskip=17pt \parindent=3ex \small
			We study the implications for  $B_s - \bar B_s$ mixing and the neutron electric dipole moment (EDM) in a light dark matter model with sizable invisible rare meson decays to accommodate the recent possible deviations from Standard Model (SM) predictions  observed in $B^+\to K^+\nu\bar\nu$ by Belle II and $K^+\to\pi^+\nu\bar\nu$ by NA62. Given that the neutrinos in these decays escape detection, they can be replaced by other invisible final states. Based on effective operator analysis, it has been proposed that branching ratios for $B^+\to K^+ +\mbox{invisible}$ and $K^+\to\pi^+ +\mbox{invisible}$ can naturally be larger than the SM predictions  due to the emission of light dark matter pairs. We demonstrate that this scenario can be realized within a UV-complete two-Higgs-doublet model (2HDM) where neutral Higgs bosons mediating dark matter interactions induce significant low-energy effects especially for $B_s-\bar B_s$ mixing and neutron EDM.  Within the allowed parameter space, we find non-negligible contributions to $B_s - \bar B_s$ mixing.  For neutron EDM, there is a cancellation due to the exchange of neutral spin-zero particle, but QCD renormalization group evolution will lift this cancellation which in fact is generally true for any neutral Higgs contribution. However, we demonstrate that such a cancellation does not occur for charged scalar contributions. Ultimately, the allowed CP-violating phases in the Yukawa sector can generate a neutron EDM at a level consistent with current bound. 
	\end{minipage}}
	
	\maketitle
	
	\newpage

	\section{Introduction\label{intro}}
	
	Rare processes serve as sensitive probes for New Physics (NP) beyond the Standard Model (SM), with $B^+\to K^+ \nu \bar \nu$ and  $K^+\to \pi^+ \nu \bar \nu$ being two prototypical examples. In the SM, the calculations of these branching ratios are relatively precise compared to decays with charged leptons or more hadrons in the final state, resulting in \,${\cal B}(B^+ \to K^+ \nu \bar \nu)_{\rm SM} = (4.43\pm 0.31)\times 10^{-6}$~\cite{Becirevic:2023aov,He:2023bnk}, where the tau-lepton-mediated tree-level contribution has been removed. The $K^+\to\pi^+\nu\bar\nu$ prediction is also remarkably precise~\cite{Brod:2010hi} but still suffers from significant parametric uncertainty arising from the input values of the Cabibbo-Kobayashi-Maskawa (CKM) angles. This is illustrated by the three different branching-ratio numbers quoted in Ref.~\cite{NA62:2024pjp}. Conservatively, adopting the value with the largest quoted uncertainty, \,${\cal B}(K^+\to \pi^+\nu\bar\nu)_{\rm SM} = (8.4\pm 1.0)\times 10^{-11}$~\cite{Buras:2015qea}, provides a baseline for discussing NP possibilities. 
	
	Recent results for these two processes have been reported by Belle II~\cite{Belle-II:2023esi} and NA62~\cite{NA62:2024pjp}, respectively. In 2023, the Belle II experiment reported a measurement of ${\cal B}(B^+ \to K^+ \nu \bar \nu) = (2.3\pm 0.7)\times 10^{-5}$~\cite{Belle-II:2023esi}, representing a $2.7\sigma$ excess over the SM prediction. Combining this with the earlier findings by BaBar~\cite{BaBar:2010oqg,BaBar:2013npw}, Belle~\cite{Belle:2013tnz,Belle:2017oht}, and Belle~II~\cite{Belle-II:2021rof} yields the weighted average \,${\cal B}(B^+ \to K^+ \nu \bar \nu)_{\rm exp} = (1.3 \pm 0.4)\times 10^{-5}$~\cite{Belle-II:2023esi}, which remains $2.1\sigma$ above the SM expectation. On the other hand, the NA62 collaboration recently reported the first observation of kaon decay $K^+\to\pi^+ + \nu \bar \nu$. The measured branching ratio is ${\cal B}(K^+\to\pi^+ + \nu\bar \nu )=13.0_{-3.0}^{+3.3} \times10^{-11}$~\cite{NA62:2024pjp}, marking it as the rarest decay mode discovered to date. While this result is statistically consistent with the SM prediction within $2\sigma$, its central value exceeds the SM expectation by $\Delta{\cal B}_K=(4.6^{+3.4}_{-3.2})\times 10^{-11}$. 
	
	Experimentally, the final-state neutrino-antineutrino pairs escape detection, meaning that the observed signals are effectively $B^+\to K^+ +\slashed{E}$ and  $K^+\to \pi^+ +\slashed E$ where $\slashed E$ denotes the missing energy carried by invisible final states. While exclusively attributed to neutrinos in the SM, this signature allows for a window of NP where light dark matter (DM) pairs mimic the missing energy. The consistently higher central values for both $B^+\to K^+ +\slashed{E}$ and  $K^+\to \pi^+ +\slashed E$ compared to SM expectations provide a compelling motivation to explore such scenarios. Previous studies have demonstrated that light DM scenarios, framed within effective Lagrangians, are phenomenologically viable in accommodating the Belle II results~\cite{He:2024iju}. Consequently, several theoretical efforts have been devoted to developing UV-complete frameworks~\cite{He:2025jfc, He:2025wxn}. Furthermore, recent effective operator analyses have been extended to incorporate the latest NA62 data~\cite{He:2025sao}. 
	
	Building upon these developments, in this work we demonstrate that the effective descriptions for both Belle II and NA62 enhanced data compared with SM predictions can be simultaneously realized within a Type-III two-Higgs-doublet model (2HDM). The addition of the second Higgs doublet introduces Yukawa couplings that can affect other processes of interest and provides new search ground for NP. In particular, in the minimal version of this model, incorporating only the minimal set of Yukawa couplings required to address the enhancement of rare decays, we study the implications for $B_s - \bar B_s$ mixing and the neutron electric dipole moment (EDM).
	Within the allowed parameter space, we find non-negligible contributions to $B_s - \bar B_s$ mixing. For neutron EDM, there is a cancellation due to the exchange of neutral spin-zero particles, but QCD renormalization group evolution will lift this cancellation which in fact is generally true for any neutral Higgs contribution. However, we demonstrate that such a cancellation does not occur for charged scalar contributions. Ultimately, the allowed CP-violating phases in the Yukawa sector can generate a neutron EDM at a level consistent with current bound. 
    \\
	\section{The minimal effective operators required}

	To accommodate potential excess in $B^+ \to K^+ +\slashed{E}$ and allow for $K^+ \to \pi^+ + \slashed{E}$ branching ratios exceeding SM prediction, it has been proposed that additional missing energy signals may arise from a light real scalar DM $\phi$, produced in pairs ($\phi\phi$) as a substitute for $\nu\bar\nu$~\cite{He:2024iju, He:2025jfc, He:2025wxn, He:2025sao, Badin:2010uh,Kamenik:2011vy}. To ensure the stability of the DM candidate, the field $\phi$ is assumed to be a SM-gauge singlet with a mass $m_\phi < (m_K - m_\pi)/2 = 177\ \mathrm{MeV}$, and is charged under some symmetry of a dark sector beyond the SM or odd under a $\mathbb{Z}_2$ symmetry.  In a Low-Energy Effective Field Theory (LEFT) framework, the lowest-dimension operators describing the $\phi$-quark couplings that can accommodate the observed data are given by~\cite{He:2024iju, He:2025jfc, He:2025wxn,He:2025sao, Badin:2010uh,Kamenik:2011vy}
	\begin{align}
		\mathcal{L}_{qq\phi^2} & =
		\frac{1}{2} \Big[ C_{d\phi}^{S,ij} \big( \overline{d_i} d_j\big) + C_{d\phi}^{P,ij} \big( \overline{d_i} i\gamma_5^{} d_j \big) 
		+ C_{u\phi}^{S,kl} (\overline u_k u_l)  \phi^2)  
		+ C_{u\phi}^{P,kl} (\overline u_k i\gamma_5 u_l) \Big] \phi^2 \,,     
		\label{eq:LEFT}
	\end{align}
	where the $C$s are generally complex, independent constants. The Hermiticity of this Lagrangian implies that \,$C_{d(u)\phi}^{S(P),ij}=C_{d(u)\phi}^{S(P),ji*}$. The minimal set of non-zero coefficients includes $(ij) = (bs), (sd), (ss), (dd)$ and $(kl) = (uu)$.  
	
	Only the $(\bar d_i d_j) \phi^2$ term in Eq.\,(\ref{eq:LEFT}) can supply the $bs\phi^2$ and $sd\phi^2$ couplings that contribute to \,$B^+\to K^+ + \slashed E$ and \,$K^+\to\pi^+ + \slashed E$,\, respectively, as the matrix element of a pseudoscalar quark bilinear between two pseudoscalar mesons vanishes. Based on $B\to K$ transition form factors and chiral perturbation theory calculations for $K \to \pi$ transitions, the required ranges of $C^{S, bs}_{d\phi}$ and $C^{S, sd}_{d\phi}$ to accommodate the observed excesses in $B^+\to K^+$+$\slashed E$ and $K^+\to\pi^+$+$\slashed E$ can be determined. These values are given by~\cite{He:2025sao}. To provide the enhanced value for $B^+\to K^+ + \slashed E$, one needs
	\begin{align}
		C^{S, bs}_{d\phi} \sim (2.5 - 6.0)\times 10^{-5}\ \mathrm{TeV}^{-1}. 
	\end{align}
	
	For $K^+\to \pi^+ + \slashed E$, we study the interesting possibility that NP accounts for the discrepancy between the SM expectation and the experimental data, where $\Delta{\cal B}_K=(4.6^{+3.4}_{-3.2})\times 10^{-11} $. This leads to 
	\begin{align}
		C^{S, sd}_{d\phi} \sim (0.06 - 0.5)\times 10^{-7}\ \mathrm{TeV}^{-1}\;,
	\end{align}
	and the $\phi$ mass is also restricted to be between $110\ \mathrm{MeV}$ and $146\ \mathrm{MeV}$.
	
	For $\phi$ to be the cosmological DM candidate, its annihilation should generate the observed relic abundance via the freeze-out mechanism. Since the $\phi$ mass must be \,$m_\phi<177\ \mathrm{MeV}$,  the relic density is primarily determined by the $(\bar uu) \phi^2$ and $(\bar dd) \phi^2$ operators. These operators contribute to $\phi \phi \to \pi\pi$ at tree level and $\phi\phi \to \gamma\gamma$ at loop level, requiring non-zero Wilson coefficients. However, these annihilation processes are simultaneously subject to stringent indirect detection constraints, particularly from Cosmic Microwave Background (CMB) observations, which limit the energy injection into the early universe. The interplay between achieving the correct relic abundance and satisfying these CMB bounds restricts the viable DM mass to a narrow window of $110\ \mathrm{MeV}-146\ \mathrm{MeV}$~\cite{He:2025sao}. Within this range, the combination of coefficients $|C^{S, uu}_{u\phi} + C^{S, dd}_{d\phi}|$ is constrained to be within $0.04-1\ \mathrm{TeV}^{-1}$~\cite{He:2025sao} . In particular, for a benchmark mass of $m_\phi=130\ \mathrm{MeV}$, this combination takes a value of $0.1\ \mathrm{TeV}^{-1}$ . 
	
	Recent results from dark matter (DM) direct-detection experiments have imposed stringent bounds on sub-GeV DM candidates, particularly when the Migdal effect is accounted for. In our scenario, the tension with these limits can be alleviated by incorporating the $(\bar ss)\phi^2$ operator, which does not affect the relic density.  The allowed parameter space can be expressed in terms of the ratios $R_s = C^{S, ss}_{d\phi} /(C^{S, uu}_{u\phi} + C^{S, dd}_{d\phi})$ and $R_- = (C^{S, uu}_{u\phi} - C^{S, dd}_{d\phi})/(C^{S, uu}_{u\phi} + C^{S, dd}_{d\phi})$. When $m_\phi=130\ \mathrm{MeV}$, these ratios are constrained to $-29<R_{-}<3$ and $-80<R_s<80$ respectively~\cite{He:2025sao}. In our subsequent analysis, we adopt a set of benchmark values centered within these allowed regions, specifically: $C^{S, ss}_{d\phi} = 0$, $C^{S,dd}_{d\phi}=-(7/6)C^{S,uu}_{u\phi} = 0.70\ \mathrm{TeV}^{-1}$. 
	
	For concreteness, we will take the benchmark points for the Wilson coefficients to be~\cite{He:2025sao}
	\begin{align}\label{wilson}
		C^{S, bs}_{d\phi} &= 6.0\times 10^{-5}\ \mathrm{TeV}^{-1} \;,\;\;\;\; &C^{S, sd}_{d\phi} &= 2.0\times 10^{-8}\ \mathrm{TeV}^{-1}\;, \nonumber\\
		C^{S, dd}_{d\phi} &= 0.7\ \mathrm{TeV}^{-1}\;,\;\;\;\;&C^{S, uu}_{u\phi} &= -0.6\ \mathrm{TeV}^{-1}\;,\nonumber\\
	\end{align}
	with the DM mass fixed at $m_\phi = 130\ \mathrm{MeV}$  which at 1$\sigma$ level can raise the branching ratio to the central measured $K^+\to \pi^+ +\slashed{E}$ value,  for the numerical analysis. We have set $C^{S, bs}_{d\phi}$ to its largest allowable value to maximize the potential effects.
	
In the following we show that a UV-complete and renormalizable framework can be constructed to realize the phenomenological features discussed above.  To this end, we explore a Type-III Two-Higgs-Doublet Model extended with a scalar dark matter field.  We will then study the model implications for $B_s - \bar B_s$ mixing and neutron EDM. Our analysis shows that these observables can satisfy current experimental limits, offering promising avenues for future experimental tests. 
	\\
	
	\section{A type-III two Higgs doublet model realization}
	
	Among the various types of two-Higgs-doublet models, we focus on the type-III 2HDM , where both Higgs doublets couple to quarks. This framework is supplemented by a real scalar $\phi$ which serves as a DM candidate. An important feature of the type-III model is that it can offer flavor-changing neutral-Higgs transitions at tree level which can readily achieve the $b\to s$ and $s\to d$ transitions for $B^+\to K^+ + \slashed E$ and $K^+\to \pi^+ + \slashed E$ with $\phi\phi$ to produce the missing energy required. The introduction of new interactions inevitably subjects the model to a wider array of constraints. While this framework has been previously explored to address the $B^+\to K^+ + \slashed E$ excess and CP violation in hyperon decays~\cite{He:2025jfc, He:2025wxn}, we extend the analysis here to investigate whether the model can simultaneously accommodate the latest results from both $B^+\to K^+ + \slashed E$ and $K^+\to \pi^+ + \slashed E$ via the $\phi\phi$ channel. We find that the model survives all known constraints and also has interesting consequences for $B_s - \bar B_s$ mixing and neutron EDM. In the following we give details. 
	
	In this framework the Higgs-quark interactions are described by the Yukawa Lagrangian~\cite{Branco:2011iw}
	\begin{align} \label{LYo}
		{\cal L}_Y  & \,=\, -\big(\hat Y_a^D  \big)_{jk}\, \overline q_j  d _k\, H_a  - \big(\hat Y_a^U \big)_{jk}\, \overline q_j   u_k  \tilde H_a\,+\, \rm H.c. \,, &
	\end{align}
	where \,$a=1,2$\, and \,$j,k=1,2,3$\, are summed over, $q_j$ $(d_k$ and $ u_k)$ represent the left-handed quark doublets (right-handed down- and up-type quark fields, respectively), $H_{1,2}$ denote the Higgs doublets, \,$\tilde H_a = i\tau_2 H_a^*$,\, with $\tau_2^{}$ being the second Pauli matrix, and $Y _a^{D,U}$ are 3$\times$3 matrices for the Yukawa couplings. In general both Higgs doublets will develop non-zero vacuum expectation values (vev). For convenience we perform a rotation on the Higgs field so that only one doublet, $H_1$, acquires a non-zero vev, while $\langle H_2\rangle=0 $. The components of the two doublets are parameterized as 
	\begin{align}
		H_1^{} & \,=\, \Bigg(\!\begin{array}{c} w^+ \\ \frac{1}{\sqrt2} \big(v^{}+ h + i z\big) \end{array}\!\Bigg) \,, &
		H_2^{} & \,=\, \Bigg(\!\begin{array}{c} H^+ \\ \frac{1}{\sqrt2} \big( H + i  A\big) \end{array}\!\Bigg) \,, &
	\end{align}
	where $v\simeq246\ \mathrm{GeV}$ is the vev providing masses for SM fields. The components $w^+$ and $z$ are the would-be goldstone bosons eaten by the gauge bosons $W$ and $Z$. $h$ and $H$ are the two physical neutral scalars and $A$ is the physical neutral pseudoscalar in the model, and $H^+$ is the physical charged scalar.  Whether they are already in their mass eigenstates depends on the parameters of the Higgs potential. The most general Higgs potential can be written as
	\begin{align}
		V = &\mu^2_1 H^\dagger_1 H_1+ \mu^2_2 H_2^\dagger H_2 + (\mu^2_{12} H^\dagger_1 H_2 + H.c.) +\kappa_1 (H^\dagger_1 H_1)^2 + \kappa_2 (H^\dagger_2 H_2)^2&\nonumber\\
		& + \kappa_3 (H^\dagger_1 H_1)(H^\dagger_2 H_2)+ \kappa_4 (H^\dagger_1 H_2)(H^\dagger_2 H_1) + \left[\kappa_5(H^\dagger_1 H_2)(H^\dagger_1 H_2) + H.c.\right]&\nonumber\\
		&+\left[(\kappa_6 H^\dagger_1 H_1 + \kappa_7H^\dagger_2 H_2)H^\dagger_1 H_2 + H.c.\right]\;.
	\end{align}
	If there is no CP violation in the potential, all parameters are real. Furthermore, if $\kappa_6=0$, the potential is in the alignment limit where $h$, $H$ and $A$ are already in their mass eigenstates. We have
	\begin{align}
		m^2_h &= 2 \kappa_1 v^2\;,\;\;\;\;&m^2_{H^\pm} &=\mu^2_2 + {1\over 2} \kappa_3 v^2,\nonumber\\
		m^2_H &= \mu_2^2 + {1\over2} (\kappa_3 + \kappa_4+ 2\kappa_5) v^2\;,\;\;\;\;&m^2_A &= \mu_2^2 + {1\over 2} (\kappa_3 + \kappa_4 - 2\kappa_5) v^2\;.
	\end{align}
	
	To satisfy the constraints from the electroweak precision observables, specifically the $S$ and $T$ parameters, we consider a scenario where $H$ and $H^\pm$ are nearly degenerate in mass and are significantly heavier than the electroweak scale. However, the pseudoscalar $A$ is allowed to remain relatively light. 
	
	In this model, $h$ identifies with the SM-like Higgs boson , responsible for the masses of the $W$ and $Z$, but $H$ and $A$ do not. It also couples to quarks and charge leptons through $Y_1$, reproducing the standard SM mass relations. We have
	\begin{align} \label{vvh}
		{\cal L}_{\rm K}^{} &
		\,=\, \frac{h}{v} \big(2m_W^2W^{+\mu}W_\mu^-+m_Z^2Z^\mu Z_\mu \big)& \nonumber\\
		{\cal L}_{h} &\,=\, -\tfrac{1}{\sqrt2} \Big[ \big (Y_1^D\big)_{jk} \overline{ D_j} P_R  D_k + \big( Y_1^U \big)_{kj}^* \overline{U_j} P_L U_k^{}      \Big] (v+h)  + H.c&
	\end{align}
	where $P_{L,R}=\big(1\mp\gamma_5^{}\big)/2$,\, and  \, $ D_{1,2,3} =(d,s,b)$ and \,$ U_{1,2,3} =(u,c,t)$. After rotating the quark fields to their mass eigenstates, the coupling of  $h$ to fermions simplifies to 
	\begin{align} \label{vvh}
		{\cal L}_{h} &\,=\, - \left[ \overline{ D} \hat M^D {} D + \overline{ U} \hat M^U  U \right]\left(1+{h\over v}\right)\;,
	\end{align}
	where $\hat M^D$ and $\hat M^U$ are the diagonal mass matrices. 
	
	The Yukawa interactions of the heavy Higgses $H$, $A$ and $H^+$ become
	\begin{align} \label{LY}
		{\cal L}_{Y}^{} & = -\, \tfrac{1}{\sqrt2} \Big[ \big( Y_2^D \big)_{jk} \overline{ D_j} P_R^{}  D_k^{} + \big( Y_2^U\big)_{kj}^* \overline{ U_j} P_L U_k \Big] (H+ iA) &\nonumber\\
		&- \overline{ U_j} \Big[ \big( V Y_2^D\big)_{jk} P_R - \big( Y_2^{U \dagger} V\big)_{jk} P_L \Big]  D_k H^+  +  H.c. \,, &
	\end{align}
	where the transformed Yukawa matrices $ Y_2^{D,U}$ are generally nondiagonal, and $V$ is the CKM matrix. One can similarly work out the lepton mass and Yukawa terms. 
	
	In the DM sector, to  ensure the stability of the DM field, $\phi$, we assume it to be a SM-gauge singlet and introduce an unbroken $Z_2$ symmetry under which \,$\phi\to-\phi$,\, while the other fields remain unaltered. The renormalizable Lagrangian for the DM sector is given by \,${\cal L}_\phi = (1/2) \partial^\mu \phi\,\partial_\mu\phi -  V_\phi$,  where~\cite{Bird:2006jd}
	\begin{align} \label{LD2hdmd}
		V_\phi & \,=\, \tfrac{1}{2} m_0^2\phi^2 + \left( \lambda_1 H_1^{\dag}H_1^{}
		+ \lambda_2^{}H^\dagger_2 H_2^{} + \lambda_3^{} H_1^{\dag}H_2^{}
		+ \lambda_3^* H_2^\dagger H_1^{} \right)  \phi^2 + \tfrac{1}{4} \lambda_\phi^{}\phi^4 \,, &\nonumber\\
		& \,=\, \tfrac{1}{2} m_0^2\phi^2 +  \left[ \lambda_1^{}{1\over 2} (v+ h)^2
		+ \lambda_2^{}{1\over 2} (H^2+A^2 + 2 H^-H^+)\right.&\nonumber\\
		& \left.+ \lambda_3^{} {1\over 2} (v+h)(  H + i A)
		+ \lambda_3^* {} {1\over 2} (v+h)(H-iA)  \right]  \phi^2 + \tfrac{1}{4} \lambda_\phi \phi^4 \,, &
	\end{align}
	with $m_0^2$ and $\lambda_{1,2,\phi}$ being real constants due to the hermiticity of ${ V}_\phi$. $\lambda_3$ can be made real by absorbing the phase into redefinition of $H_2$ which effectively eliminates the $A\phi\phi$ couplings. After electroweak symmetry breaking, ${\cal L}_\phi$ contains the dark matter mass which is modified to $m_\phi^2 = m_0^2 + \lambda_1 v^2$. 
	
	The scalar potential further contains various self-interactions among the Higgs doublet components; however, these couplings do not directly impact our subsequent analysis. We assume that all such parameters are chosen to be consistent with the requirements of perturbativity, unitarity, and vacuum stability, as well as other relevant theoretical and experimental constraints~\cite{Branco:2011iw,He:2016mls}.
	
	By exchanging the neutral scalars $h$ and $H$ between quarks and $\phi$ , we obtain the effective interactions required to induce $B^+\to K^+ \phi \phi$ and $K^+\to \pi^+ \phi \phi$ decays. The effective Lagrangian is given by
	\begin{align}\label{quark-DM}
		{\cal L}^{\phi\phi}_{eff} &= \lambda_1{1\over m^2_h} \left(\bar D M^D D + \bar U M^U U\right) \phi\phi&\nonumber\\
		&+\lambda_3{v\over m^2_H}\frac{1}{2\sqrt{2}} \left(\bar D \left[Y^D_2 + Y^{D\dagger}_2 + (Y^D_2 - Y^{D\dagger}_2)\gamma_5 \right] D 
		+\bar U \left[Y^U_2 + Y^{U\dagger}_2 + (Y^U_2 - Y^{U\dagger}_2 )\gamma_5\right] U\right) \phi \phi\; ,
	\end{align}	
	from which we derive the corresponding Wilson coefficients 
	\begin{align} 
		&C^{S, ij}_{d\phi, u\phi} = 2 \lambda_1 {M^{D, U}_{ij}\over m^2_h} + \lambda_{3}  { v\over \sqrt{2} m^2_H} \left(Y^{D, U}_2 + Y^{D,U\dagger}_2\right)\;,&\nonumber\\
		&C^{P, ij}_{d\phi, u\phi} =  -i\lambda_{3}{ v\over \sqrt{2} m^2_H} \left(Y^{D, U}_2 - Y^{D,U\dagger}_2\right)\;.&
	\end{align}
	
	In addition to these DM-sector interactions, the exchange of $h$, $H$ and $A$ at tree and loop levels will also generate operators involving just SM particles and induce detectable new  effects. After the required Wilson coefficients in Eq. (\ref{wilson})  are determined, the Yukawa couplings are constrained. One then needs to examine whether the induced new effects have ruled out the model. We will study this in the next section.
	\\
	
	\section{constraints on the minimal model parameters}
	
	To reproduce the benchmark values for the observed excesses, only a specific subset of the parameters in $Y_2^{D,U}$ needs to be non-zero. The minimal requirement is that $Y_2^{D,U}$ are real, with only $Y_2^{D,bs}$, $Y_2^{D,sd}$, $Y_2^{D,dd}$, and $Y_2^{U,uu}$ being non-zero.  Under these assumptions, the relevant Wilson coefficients simplify to
	\begin{align}
		&C^{D, bs}_{d\phi} = \lambda
		_{3} {v\over {\sqrt{2}m^2_H}}Y^{D, bs}_2\;, \;\;C^{D, sd}_{d\phi} = \lambda
		_{3} {v\over \sqrt{2}m^2_H}Y^{D, sd}_2\;,&\nonumber\\
		& \;\;C^{D, dd}_{d\phi} = \sqrt{2}\lambda
		_{3} {v\over m^2_H}Y^{D, dd}_2\;, \;\;C^{U, uu}_{d\phi} = \sqrt{2}\lambda
		_{3} {v\over m^2_H}Y^{U, uu}_2\;.
	\end{align}
	Note that one could also choose $Y^{D, sb, ds}_2$ to be non-zero. As we are demonstrating the model can survive the scrutiny of data, one can take just one of them for illustration. Using the previously determined Wilson coefficients, we obtain the required Yukawa couplings are  
	\begin{align}\label{yukawa}
		&	|Y^{D,bs}_2| \approx 3.4\times 10^{-4}{1\over |\lambda_3|}\left ({m_H\over \mathrm{TeV}}\right )^2\;,  \;\;|Y^{D, sd}_2| \approx  1.1 \times 10^{-7}{1\over |\lambda_3|}\left ({m_H\over \mathrm{TeV}}\right )^2 \;,&\nonumber\\
		&
		\;\;|Y^{D, dd}_2| \approx  2.0  {1\over |\lambda_3|}\left ({m_H\over \mathrm{TeV}}\right )^2 \;,	\;\;|Y^{U, uu}_2| \approx 1.7 {1\over |\lambda_3|}\left ({m_H\over \mathrm{TeV}}\right )^2\;.
	\end{align} 

    In estimating $Y^{D, dd}_2$ and $Y^{U, uu}_2$, we have neglected small contributions from the $\lambda_1$ which is a valid approximation provided that $\lambda_1$ is not several orders of magnitude larger than $\lambda_3$. In the above, all Yukawa couplings depend on $m_H^2/|\lambda_3|$. For phenomenological studies of contributions to low energy processes, the values of $\lambda_{3}$ and $m_{H}$ should be neither too large nor too small. We constrain them from the largest Yukawa $|Y_2^{D, dd}|$ to be less than a generic perturbative bound of $\sqrt{4 \pi}$ which is used in the literature~\cite{Chakrabarty:2014aya}. Recently in Ref.~\cite{Allwicher:2021rtd}, a more stringent value of $\sqrt{8\pi/3}$ was obtained. We will use these two values as our reference benchmark points for later discussions. This then implies that $m_H$ and $|\lambda_3|$ are correlated. For example with $|Y^{D,dd}_{2}|<\sqrt{4 \pi}\  (\sqrt{8\pi/3})$, $m_H = 0.5\ \mathrm{TeV}$, $|\lambda_3|$ should be larger than $0.14\ (0.17)$, and for $m_H =1\ \mathrm{TeV}$, $|\lambda_3|$ should be larger than $0.56\ (0.69)$. We will keep these in mind for subsequent numerical discussions.

	To make sure that the $H$ Higgs mass is large enough to survive the LHC data, we will take $m_H = 0.5\ \mathrm{TeV}$ for example. We also take $H^+$ mass to be approximately degenerate with H so as not to cause potential problems for $S$, $T$ and $U$ electroweak precision observables.  We will take all non-zero $Y_2^{D, U}$ elements to be real as our working assumption in the study of $B_s- \bar B_s$ mixing and will allow them to be complex when studying the neutron EDM.
	
	With the above minimal non-zero $Y^{D, U}_2$,  Eq.~(\ref{quark-DM}) becomes
	\begin{align}\label{dm-phiphi}
		{\cal L}^{\phi\phi}_{eff} &=\lambda_{3} {v\over \sqrt{2} m^2_H}\left(Y^{D,bs}_2( \bar b P_{R} s + \bar s P_{L} b )+  Y^{D,sd}_2 (\bar s P_{R} d + \bar d P_{L} s) + Y^{D, dd}_2 \bar d  d  + Y^{U, uu}_2 \bar u u \right)\phi\phi\;.
	\end{align}
	
	Eq.~(\ref{LY}) of $H$, $A$ and $H^+$ Yukawa interactions becomes
	\begin{align}
		{\cal L}_{Y} &= -{1 \over \sqrt{2}}(Y^{D,bs}_2 \bar b P_{R} s +  Y^{D,sd}_2  \bar s P_{R} d + Y^{D, dd}_2 \bar d P_R d  + Y^{U, uu}_2 \bar u P_Lu )(H+i A) + H.c&\nonumber\\
		&-\left[V_{jb}Y^{D, bs}_2\bar u_j P_R s + V_{js}Y^{D, sd}_2 \bar u_j P_Rd + V_{jd}Y^{D, dd}_2\bar u_j P_R d - Y^{U,uu}_2 V_{uk}\bar u P_L d_k\right] H^+ + H.c.\;.
	\end{align}
	
	The exchange of the neutral scalars $H$, $A$ and the charged scalar $H^+$ generates the following four-quark effective Lagrangian
	\begin{align}\label{eq-mixing}
		{\cal L}_{eff} &= {1\over 4 m^2_H}\left(Y^{D,bs}_2( \bar b P_{R} s + \bar s P_{L} b )+  Y^{D,sd}_2 ( \bar s P_{R} d + \bar d P_{L} s) + Y^{D, dd}_2 \bar d  d  + Y^{U, uu}_2 \bar u u \right)^2&\nonumber\\
		&-{1\over 4 m^2_A}\left(Y^{D,bs}_2( \bar b P_{R} s - \bar s P_{L} b)+  Y^{D,sd}_2 (  \bar s P_{R} d - \bar d P_{L} s) + Y^{D, dd}_2 \bar d \gamma_5 d  + Y^{U, uu}_2 \bar u\gamma_5 u \right)^2&\nonumber\\
		&+{1\over m^2_{H^+}}\left[V_{jb}Y^{D, bs}_2\bar u_j P_R s + V_{js}Y^{D, sd}_2 \bar u_j P_Rd + V_{jd}Y^{D, dd}_2\bar u_j P_R d - Y^{U,uu}_2 V_{uk}\bar u P_L d_k\right]&\nonumber\\
		&\times \left[V^*_{jb}Y^{D, bs}_2\bar s  P_L u_j + V^*_{js}Y^{D, sd}_2 \bar d P_L u_j + V^*_{jd}Y^{D, dd}_2\bar d P_L u_j - Y^{U,uu}_2 V^*_{uk}\bar d_k P_R u\right] 
	\end{align}
	Using the minimal parameter set identified previously, we can satisfy the required benchmark values shown in Eq.~\eqref{yukawa}. This parameter space has been verified to remain consistent with current experimental constraints. The new scalar particles $H$, $A$ and $H^+$ can modify SM predictions for several processes, such as $B_s - \bar B_s$ mixing and neutron EDM which we study in the next section.
	\\
	
	\section{Implications for $B_s-\bar B_s$ mixing and neutron EDM}
	
	\subsection{Effects on $B_s-\bar B_s$ mixing}
	
	The minimal parameter set introduced in the previous section leads to new physics contributions to $B_s -\bar B_s$ and $K^0 - \bar K^0$ mixing. If $H$ and $A$ are degenerate, their contributions to the meson mixing amplitudes cancel each other out. A mass splitting between $m_A$ and $m_H$ is required to yield non-trivial contributions. The details are as follows.

	The operator contribution from Eq. (\ref{eq-mixing}) for $B_s -\bar B_s$ mixing is as follows
	\begin{equation}
		\mathcal{L}_{B_s -\bar B_s}=\eta{{Y^{D,bs}}^2\over 4}\left({1\over m_{H}^2} - {1\over m^2_A}\right)(\bar{b}P_{R}s) (\bar{b}P_{R}s),
	\end{equation}
	where~\cite{Buras:2000if} $\eta = \frac{1}{2}\left[ b_-\left(\frac{\alpha_s(m_Z)}
	{\alpha_s(m_b)}\right)^{3a_+/46}+b_+\left(\frac{\alpha_s(m_Z)}
	{\alpha_s(m_b)}\right)^{3a_-/46}\right]$ is the QCD running factor running from $m_Z$ down to $m_b$ with $a_\pm=(2/3)(1\pm\sqrt{241})$ and $b_\pm= (1\pm 16/\sqrt{241})$.    We use $\alpha_s(m_Z) = 0.1180$~\cite{ParticleDataGroup:2024cfk}. At the leading order we have $\alpha_s(m_b) = 0.2121$ with $m_b = 4.183\ \mathrm{GeV}$~\cite{ParticleDataGroup:2024cfk}. And we can derive $\eta\approx 1.46$.
	Using the matrix element~\cite{Lenz:2006hd}
	\begin{equation}
		\left\langle\bar{B}_s\right|\left(\bar{b} P_R s\right)\left(\bar{b} P_R s\right)\left|B_s\right\rangle=-\frac{5}{12}\frac{f_{B_s}^2 m_{B_s}^4}{\left(m_s+m_b\right)^2} B_S =-0.838\;\mathrm{GeV}^4.
	\end{equation}
	where $B_{S}=0.835$~\cite{DiLuzio:2019jyq}, $f_{B_s}=230.3\;\mathrm{MeV}$~\cite{FlavourLatticeAveragingGroupFLAG:2024oxs}, $m_{b}=4.18\;\mathrm{GeV}$, $m_{s}=93\;\mathrm{MeV}$, $m_{B_s}=5.36691\ \mathrm{GeV}$~\cite{ParticleDataGroup:2024cfk} . We obtain the mixing mass parameter $\Delta m_s^{NP}$ as
	\begin{equation}
		\Delta m_s^{NP}= 2M_{12}=-2 \times\frac{1}{2m_{B_s}}\left\langle\bar{B}_s\right|
		\mathcal{L}_{B_s -\bar B_s}\left|B_s\right\rangle=\eta \frac{5}{48}\frac{(Y^{D, bs}_2)^2}{m_H^2}\left(1-\frac{m_H^2}{m_A^2}\right)\frac{f_{B_s}^2 m_{B_s}^3}{\left(m_s+m_b\right)^2} B_S.
	\end{equation}
	Since $Y^{D,bs}_2$ is a real number,  using Eq. \eqref{yukawa}, we obtain
	\begin{equation}
		\Delta m_s^{NP} =2.58\times10^{-3}
		|Y^{D,dd}_2|^2\left(\frac{\mathrm{TeV}}{m_H}\right)^2
		\left(1-\frac{m_{H}^2}{m_{A}^2}\right)\ \mathrm{ps}^{-1} .
	\end{equation}

	The experimental value is~\cite{ParticleDataGroup:2024cfk}
	\begin{equation}
		\Delta m_s^{\mathrm{exp}} = 17.765 \pm 0.004 (\mathrm{stat}) \pm 0.004(\mathrm{syst})\  \mathrm{ps}^{-1}.
	\end{equation}
	
\begin{figure}[t]
		\centering
		\includegraphics[scale=0.5]{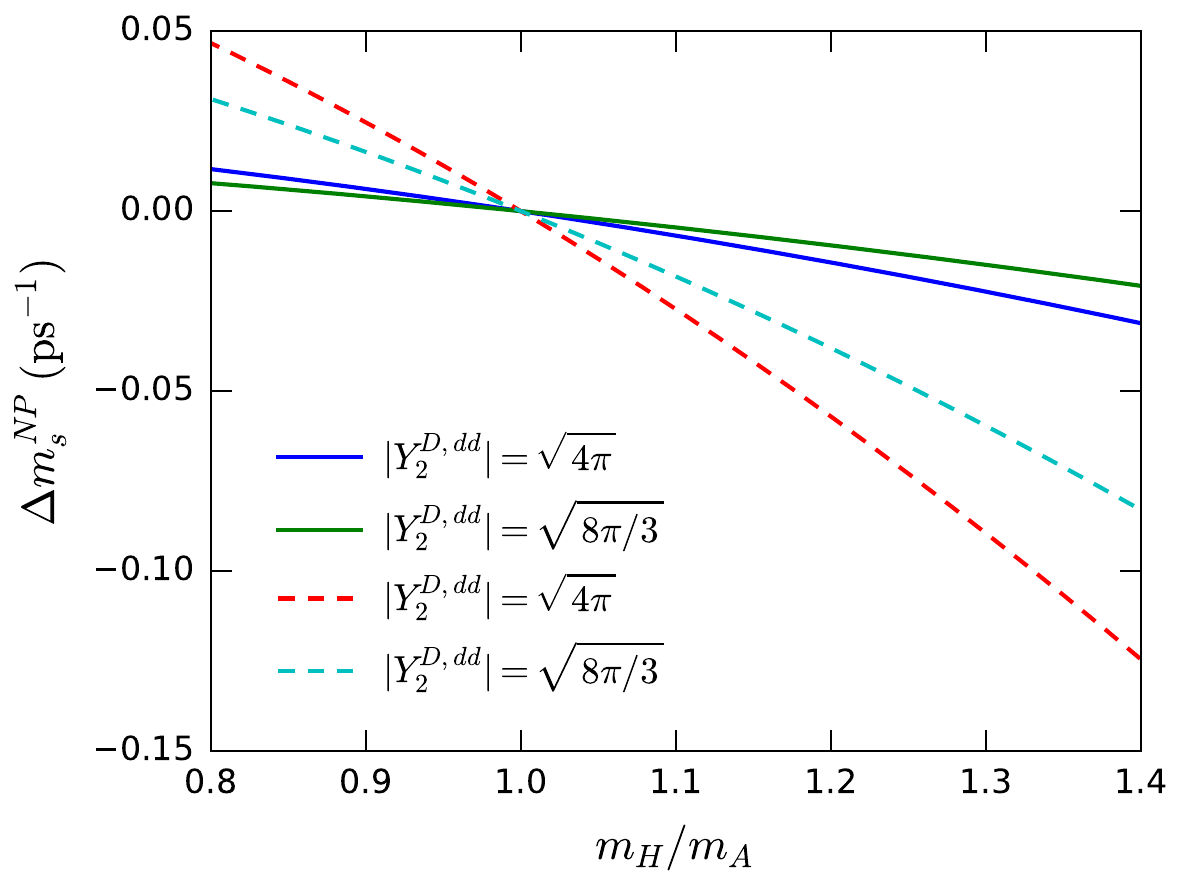}
		\caption{$\Delta m_s^{NP}$ as a function of $m_H/m_A$ at different $|Y_2^{D, dd}|$. The solid line represents $m_H=1\ \mathrm{TeV}$ (the horizontal axisranges from 0.94 to 1.07 for $|\kappa_5| <1$). The dashed line represents $m_H=0.5\ \mathrm{TeV}$ (the horizontal axis ranges from 0.8 to 1.4). }
		
		\label{delta_ms m_A figure}
	\end{figure}

	Before comparing with data, one should subtract the SM contribution which has been evaluated~\cite{Albrecht:2024oyn,UTfit:2022hsi}. We will use $\Delta m_s^{SM} = (18.23 \pm 0.63)\  \mathrm{ps}^{-1}$~\cite{Albrecht:2024oyn} for discussion. This implies that the NP contribution is constrained to the interval $\Delta m_s^{NP}=-0.465\pm 0.63\ \mathrm{ps}^{-1}$.  It is noteworthy that the SM prediction is slightly larger than the observed value. If this discrepancy is attributed to new physics, our model would require $m_A<m_H$ to provide a negative contribution to the mass difference. Our results are shown in Fig. \ref{delta_ms m_A figure}.
	
	Note that $m_H^2-m_A^2=2\kappa_5v^2$ where $v=0.246\ \mathrm{TeV}$, so when $m_H$ is fixed, the range of $m_A$ is only determined by $\kappa_5$. We take $|\kappa_5|\leq1$, and we can get the range of $m_H/m_A$ is about $(0.8,1.4)$ when $m_H=0.5\ \mathrm{TeV}$ and $(0.94,1.07)$ when $m_H=1.0\ \mathrm{TeV}$. In Fig.~\ref{delta_ms m_A figure}, the horizontal axis spans from $0.8$ to $1.4$. For the larger range of $m_H/m_A$, as $m_H/m_A$ increases, $\Delta m_s^{NP}$ will decrease. We can find that all the curves intersect at the same point. This is because when $m_H = m_A$ , the contribution from $H$ and $A$ particle cancel each other out and $\Delta m_s^{NP}=0\ \mathrm{ps}^{-1}$. For $m_H\neq m_A$, the NP contribution to $B_s - \bar B_s$ mixing falls within the expected range, namely $\Delta M_s^{NP} = -0.465 \pm 0.63\ \mathrm{ps}^{-1}$. As illustrated in Fig.~\ref{delta_ms m_A figure}, the NP contribution can reach a maximum size as low as about $-0.13\ (-0.08)\ \mathrm{ps}^{-1}$ for $|Y^{D,dd}_2|=\sqrt{4\pi}\ (\sqrt{8\pi/3})$, which is significant compared to $\Delta M_s^{NP}=-0.465\pm 0.63\ \mathrm{ps}^{-1}$. If more precise SM and experimental values in the future narrow down the allowed region in Fig.~\ref{delta_ms m_A figure}, constraints may be imposed on the maximal allowed $Y_2^{D, dd}$. 
		
	In the limit $m_H/m_A$ larger than 1, for the Wilson coefficients used in Eq.~(\ref{wilson}), the parameters chosen for the model satisfy other rare $B$ and $B_s$ decay constraints, as can be inferred from the results in Ref.~\cite{He:2024iju}. We now consider $B_s \to \phi\phi$, $B \to X_s \gamma$,  $b\to s u\bar u$ and $b\to s d \bar d$ due to the Lagrangian in Eqs.~(\ref{dm-phiphi}) and (\ref{eq-mixing}).  The constraint from $B_s\to \phi\phi$ comes from 
	$\mathcal{B}(B_s \to \mathrm{invisible}) < 5.6\times 10^{-4}$. It has already been pointed out~\cite{He:2024iju} that the experimental bound is satisfied when taking the benchmark value $C^{S, bs}_{d\phi} =  6 \times 10^{-5}\ \mathrm{TeV}^{-1}$ as specified in Eq.~\eqref{wilson}. For $B \to X_s \gamma$, the contribution is negligibly small because in this minimal model it is proportional to $C^{S,bs}_{d\phi}$ which is only $6 \times 10^{-5}\ \mathrm{TeV}$.  For 
	$b\to s (u \bar u, d \bar d)$, the parameterized amplitudes are given as $A_{q} \bar s P_L b \bar q q$. From Eq.~\eqref{eq-mixing}, we have $A_u = Y^{D, bs}_{d\phi}Y^{U, uu}_{u\phi}/(2m_H^2)\sim 7.2\times10^{-11}(Y^{D,dd}_{2})^2\ \mathrm{GeV}^{-2}$ and $A_d = Y^{D, bs}_{d\phi}Y^{D, dd}_{d\phi}/(2m_H^2)\sim  8.5\times10^{-11}(Y^{D,dd}_{2})^2\ \mathrm{GeV}^{-2}$. While in the SM, these decays are due to $\Delta S = 1$ penguin contributions, with the equivalent coefficients $A(q= u, d)$ being of order  $1.3\times 10^{-8}$. For $|Y^{D,dd}|$ equals to its perturbative limit of $\sqrt{4\pi}$ or $\sqrt{8\pi/3}$, the new contribution will still not spoil the SM predictions.
	
	The exchange of $H$ and $A$ also contributes to $K^0 - \bar K^0$ mixing through the same mechanism as in the $B_s$ system. Given that $|Y^{D, sd}_2| \sim 10^{-4} |Y^{D, bs}_2|$, the NP contribution to the mass difference $\Delta m_K^{NP}$ can be estimated as $\Delta m_{K}^{NP}\approx10^{-10}(Y^{D,dd}_2)^2\left(\mathrm{TeV}/m_H\right)^2
		\left(1-m_{H}^2/m_{A}^2\right)$. This gives $\Delta m_K^{NP} \sim \mathcal{O}(10^{-9})\ \mathrm{ ps}^{-1}$, which is roughly six orders of magnitude smaller than the experimental value $\Delta m_{K}^{\mathrm{exp}}=(5.293\pm0.009)\times
	10^{-3}\ \mathrm{ps}^{-1}$. Therefore, Kaon mixing does not impose further constraints on the parameter space considered here. 
	\\
	
	\subsection{ Effects on neutron EDM}
	
	We now discuss the potential impact on the CP-violating neutron EDM. In the SM, the neutron EDM is predicted to be several orders of magnitude~\cite{McKellar:1987tf, He:1989mbz} smaller than the current experimental upper bound of $|d_n| <1.8\times10^{-26}\ e\cdot\mathrm{cm}$ ($90\%$ C.L.)~\cite{Abel:2020pzs}. If all Yukawa couplings in Eq.~(\ref{yukawa}) are real, no new CP-violating source is introduced to generate a non-zero neutron EDM. The current constraints on the first-generation Yukawa couplings in our model are derived from the Migdal effect, which primarily probes the real parts of these couplings. However, since the magnitudes of the $u$- and $d$-quark Yukawa couplings in our framework are significantly larger than those in the SM, one-loop effects could become substantial if complex phases are allowed, i.e., $Y^{D, dd}_2 \to |Y^{d, dd}_2|e^{i\theta^{dd}}$ and $Y^{D, uu}_2 \to |Y^{U,uu}_2|e^{i\theta^{uu}}$. In such a scenario, the values in Eq.~(\ref{yukawa}) should be interpreted as $Y^{qq}_2 \cos\theta^{qq}$. If these phases $\theta_{qq}$ are non-zero, they could lead to sizable contributions to the neutron EDM. 
	
	With $Y^{D, dd}_2$ complex, $Y^{D, dd}_2=|Y^{D, dd}_2|e^{i\theta^{dd}}$, we can get a non-zero d-quark EDM at one loop level with 
	\begin{equation}
		d_{dEDM}^{\mathrm{n}}=-\frac{Q_de|Y^{D, dd}_2|^2c_{\theta^{dd}} s_{\theta^{dd}} m_d}{(4\pi)^2m_H^2}y_d(m_H , m_A),
	\end{equation}
	where $Q_d = -\frac{1}{3},\ m_d = 4.70\ \mathrm{MeV}$~\cite{ParticleDataGroup:2024cfk}, $c_{\theta^{dd}} = \cos\theta^{dd}$ and $s_{\theta^{dd}}= \sin\theta^{dd}$.
	
	Similarly one can obtain a non-zero u-quark EDM as 
	\begin{equation}
		\begin{aligned}
			d_{uEDM}^{\mathrm{n}}&=\frac{Q_ue|Y^{U, uu}_2|^2c_{\theta^{uu}}s_{\theta^{uu}} m_u}{(4\pi)^2m_H^2}y_u(m_H,m_A),
		\end{aligned}
	\end{equation}
	where $Q_u=\frac{2}{3},m_u=2.16\ \mathrm{MeV}$~\cite{ParticleDataGroup:2024cfk}. And $y_q(m_H,m_A)$ is 
	\begin{equation}
		y_q(m_H,m_A)\equiv \left[\frac{3}{2}\left(1-\frac{m_H^2}{m_A^2}\right)+\mathrm{ln}\left(\frac{m_q^2}{m_H^2}\right)-\frac{m_H^2}{m_A^2}\mathrm{ln}\left(\frac{m_q^2}{m_A^2}\right)\right].
	\end{equation}
	In the valence quark model, the neutron EDM due to neutral Higgs exchange is
	\begin{eqnarray}
		d^{\mathrm{n}}_{nEDM} = {4\over 3} d^{\mathrm{n}}_{dEDM} - {1\over 3}d^{\mathrm{n}}_{uEDM}\;.
	\end{eqnarray}
	The same $H$ and $A$ will also generate CP violating quark chromo-dipole moment (CDM) defined by
	\begin{equation}
		\mathcal{L}_g=-g_s d_{qCDM} \frac{i}{2}\bar q\sigma_{\mu\nu}\gamma_5 T^{A} q G^{\mu\nu}.
	\end{equation}
	The one-loop $H$ and $A$ contributions to $d^{\mathrm{n}}_{qCDM}$ are given by
	\begin{equation}
		d_{dCDM}^{\mathrm{n}}=-\frac{|Y^{D,dd}_2|^2 c_{\theta^{dd}} s_{\theta^{dd}} m_d}{(4\pi)^2m_H^2}y_d(m_H,m_A).
	\end{equation}
	\begin{equation}
		\begin{aligned}
			d_{uCDM}^{\mathrm{n}}&=\frac{|Y^{U, uu}|^2c_{\theta^{uu}} s_{\theta^{uu}} m_u}{(4\pi)^2m_H^2}y_u(m_H,m_A).\\
		\end{aligned}
	\end{equation}
	In the valence quark model, the color dipole contribution to the neutron EDM $d_{qCDM}$ is given by  \cite{He:1989mbz} 
	\begin{equation}
		d_{nCDM}^{\mathrm{n}} = e\left(\frac{4}{9}d_{dCDM}^{\mathrm{n}}+\frac{2}{9}d_{uCDM}^{\mathrm{n}}\right),
	\end{equation}
	which is equal to $-d_{nEDM}^{\mathrm{n}}$. Therefore, a significant cancellation occurs between the quark EDM and CDM contributions. It is worth noting that this cancellation is not unique to our specific framework; rather, it is a general feature of models where the CP-violating effects are mediated by neutral Higgs bosons. 
	
	We would like to point out that this cancellation is at the $H$ and $A$ mass scale. When running from high energy $\mu_{h}$ down to low energy $\mu_l$, the EDM and CDM operators run differently and the cancellation is lifted as
	\begin{equation}
		\begin{aligned}
			d_{nEDM}(\mu_l)&=\eta_{EDM}d_{nEDM}(\mu_h)+\frac{8}{3}(\eta_{CDM}-\eta_{EDM})d_{nCDM}(\mu_h)\\
			d_{nCDM}(\mu_l)&=\eta_{CDM}d_{nCDM}(\mu_h)
		\end{aligned}
	\end{equation}
	where $\mu_l=1\ \mathrm{GeV}$ and
	\begin{equation}
		\begin{aligned}
			\eta_{EDM} &= \left(\frac{\alpha_s(m_Z)}{\alpha(m_b)}\right)^{\frac{16}{23}}\left(\frac{\alpha_s(m_b)}{\alpha(m_c)}\right)^{\frac{16}{25}}\left(\frac{\alpha_s(m_c)}{\alpha(\mu_l)}\right)^{\frac{16}{27}}\approx0.478\\
			\eta_{CDM} &= \left(\frac{\alpha_s(m_Z)}{\alpha(m_b)}\right)^{\frac{14}{23}}\left(\frac{\alpha_s(m_b)}{\alpha(m_c)}\right)^{\frac{14}{25}}\left(\frac{\alpha_s(m_c)}{\alpha(\mu_l)}\right)^{\frac{14}{27}}\approx0.524
		\end{aligned}
	\end{equation}
	We use $\alpha_s(m_Z) = 0.1180$~\cite{ParticleDataGroup:2024cfk} and derive the values of $\alpha_s(\mu)$ using $\alpha_s(\mu)=\alpha_s(\Lambda)/(1-\frac{\beta_0}{2\pi} \alpha
	_s(\Lambda)\mathrm{ln}(\Lambda/\mu))$, where $\beta_0 = (33-2n_f)/3$, $n_f$ is the number of flavour. 
	
	The total contribution to the neutron EDM from the exchange of neutral scalars $H$ and $A$ is given by 
	\begin{equation}\label{nEDM neutral}
		\begin{aligned}
			d_n^{}(\mu_l) &= d^{\mathrm{n}}_{nEDM}(\mu_l)+d^{\mathrm{n}}_{nCDM}(\mu_l) \\
			&=\frac{11}{3}(\eta_{EDM}-\eta_{CDM})d^{\mathrm{n}}_{nEDM}(\mu_h) \approx - 0.17 d^{\mathrm{n}}_{nEDM}(\mu_h).
		\end{aligned}
	\end{equation}
	\begin{figure}[t]
		\centering
		\begin{subfigure}{0.47\textwidth}
			\centering
			\includegraphics[width = \textwidth]{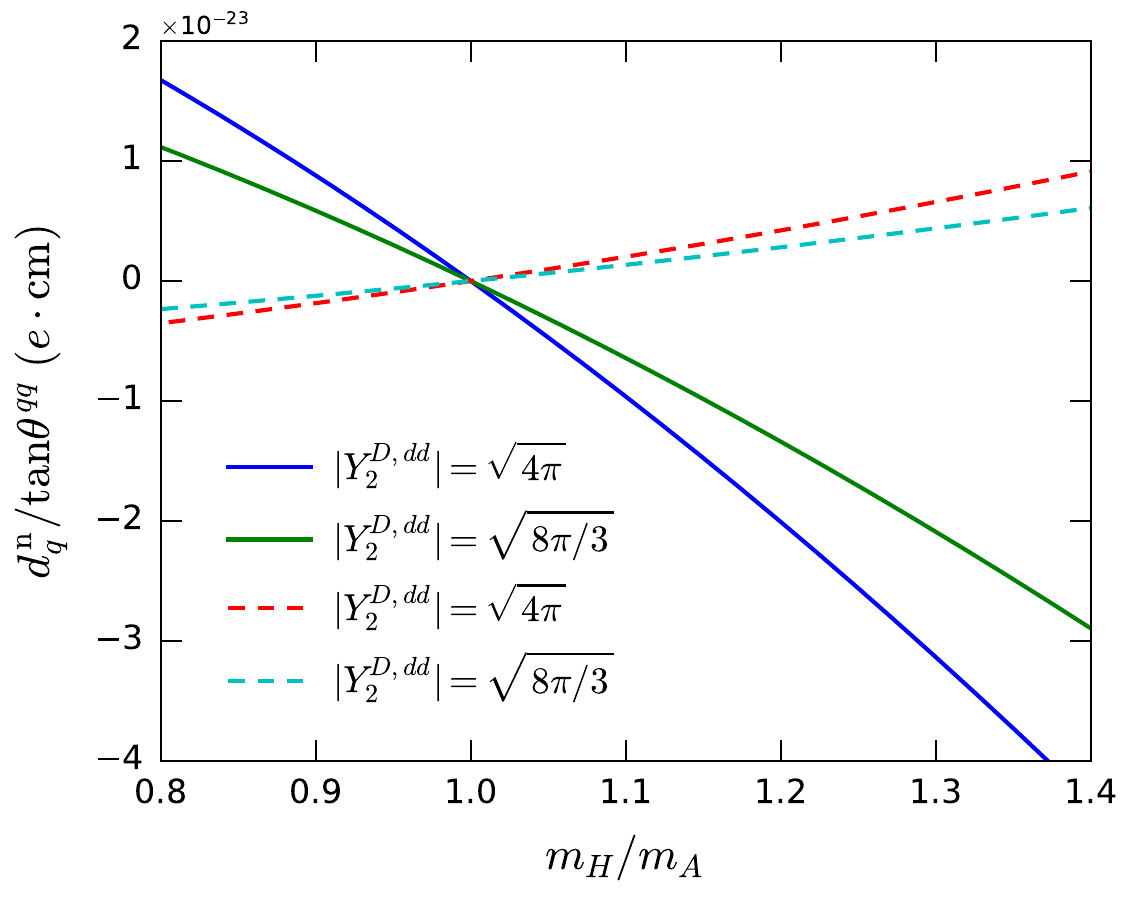}
		\end{subfigure}
		\hfill
		\begin{subfigure}{0.47\textwidth}
			\centering
			\includegraphics[width = \textwidth]{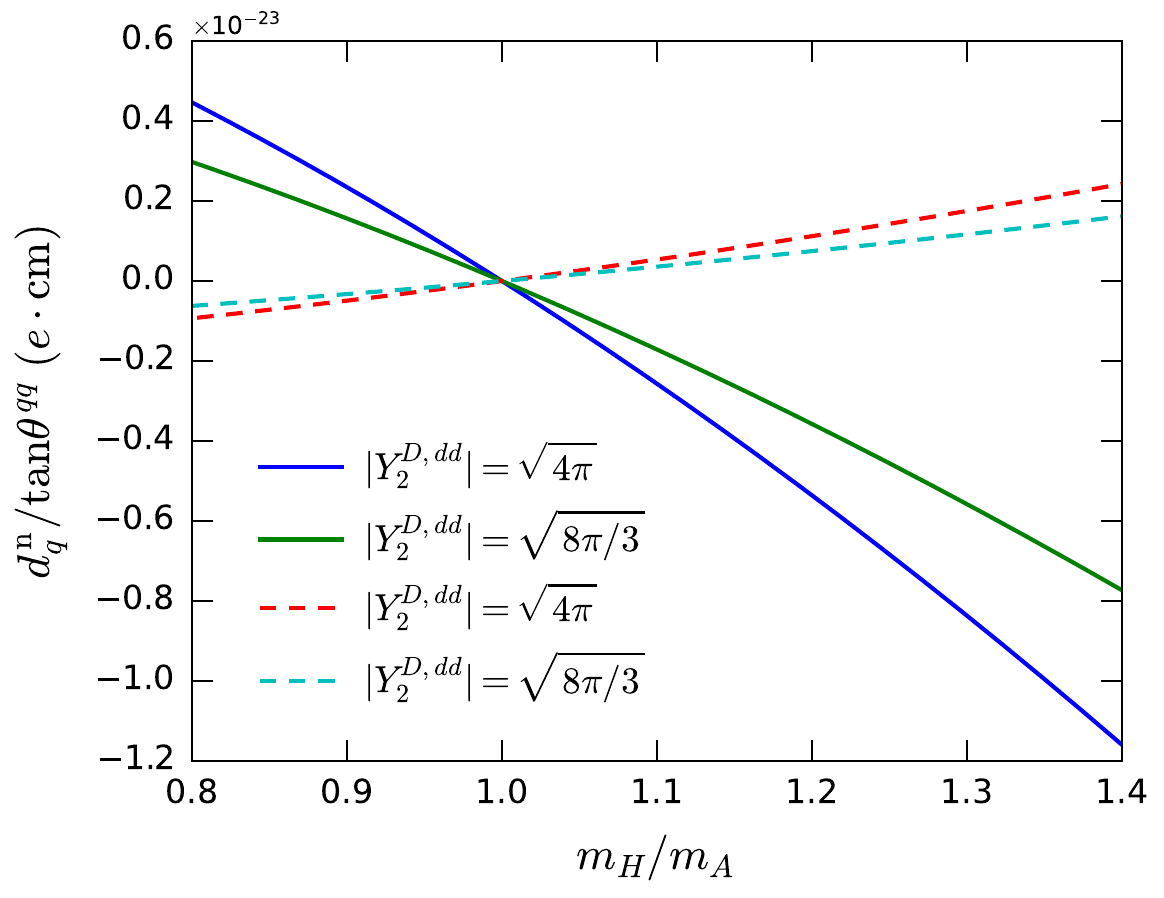}
		\end{subfigure}

		\caption{The contribution of u and d quark to neutron EDM due to the neutral Higgs divided by $\mathrm{tan}\theta^{qq}$ as a function of $m_H/m_A$ for $|Y_2^{D, dd}|$ equal to $\sqrt{8\pi/3}$ and $\sqrt{4\pi}$ respectively. The solid lines represent the d quark contribution. The dashed lines represent the u quark contribution. Left panel: $m_H=0.5\ \mathrm{TeV}$. Right panel: $m_H=1\ \mathrm{TeV}$.}
		\label{quarkEDM neutral mA}
	\end{figure}
    
	\begin{figure}[h]
		\centering
		\begin{subfigure}{0.47\textwidth}
			\centering
			\includegraphics[width = \textwidth]{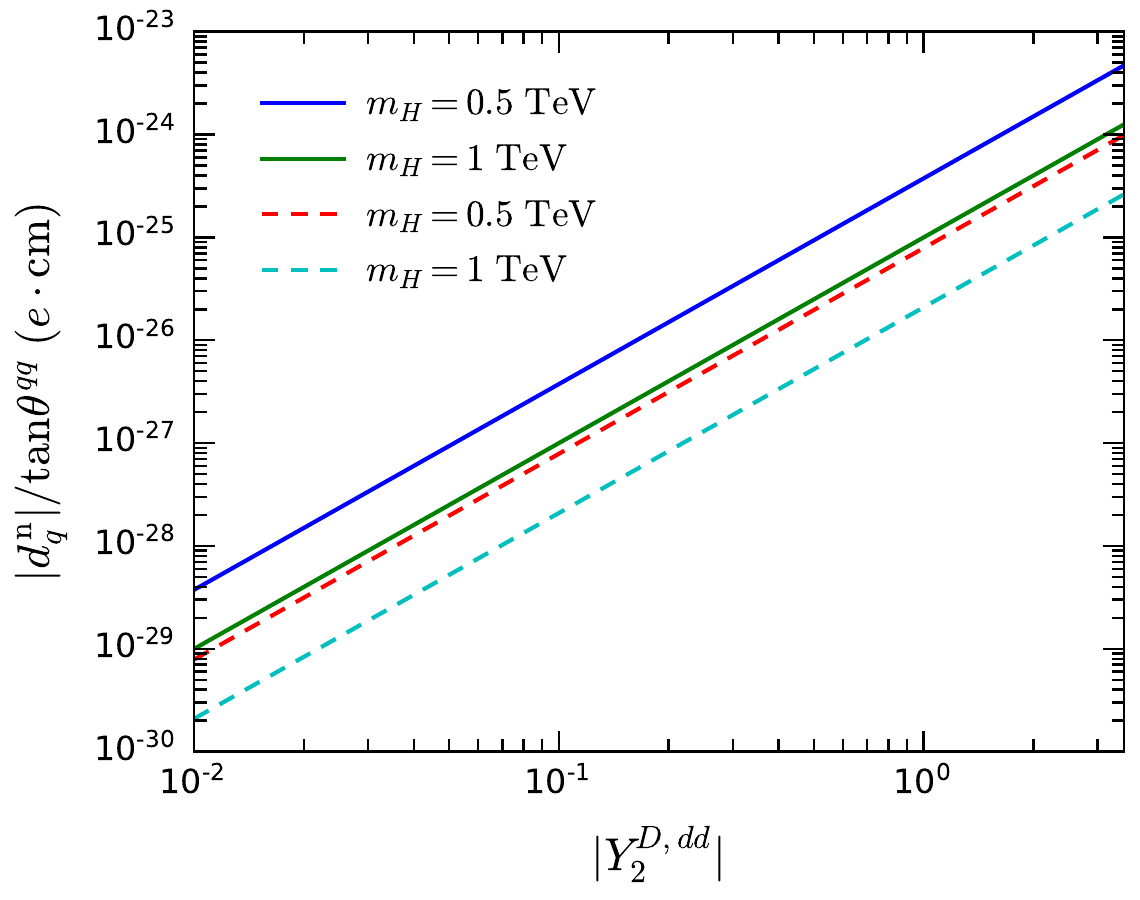}
		\end{subfigure}
		\hfill
		\begin{subfigure}{0.47\textwidth}
			\centering
			\includegraphics[width = \textwidth]{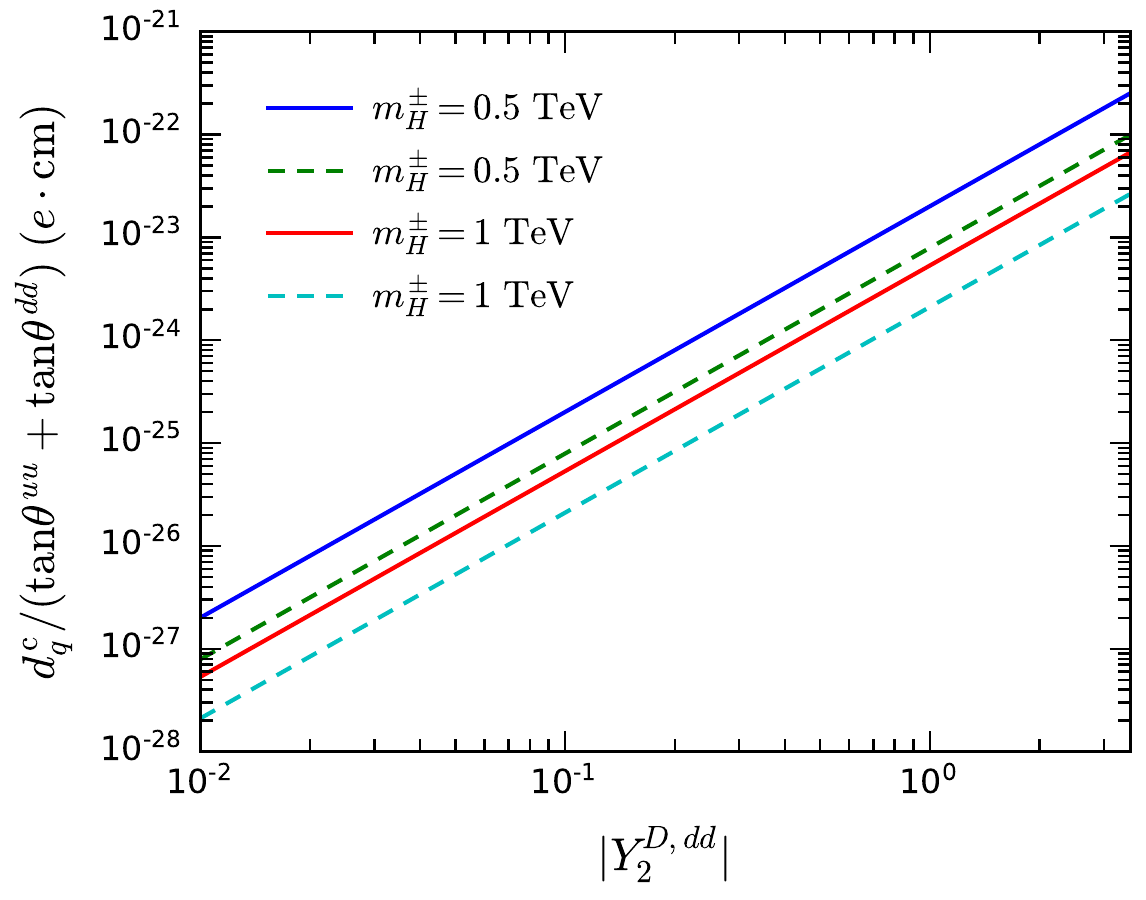}
		\end{subfigure}
		\caption{Left panel: The contribution of u and d quark to neutron EDM due to the neutral Higgs divided by $\mathrm{tan}\theta^{qq}$ as a function of $|Y_2^{D,dd}|$ with $m_H/m_A = 1.05$ . Right panel : The contribution of u and d quark to neutron EDM due to the charged Higgs divided by $\mathrm{tan}\theta^{qq}$ as a function of  $|Y_{2}^{D,dd}|$. The solid lines represent d quark contribution and the dashed lines represent u quark contribution.  }
		\label{quarkEDM lambda3}
	\end{figure}

	In the model we are studying, there are also charge Higgs contributions. The leading Yukawa coupling is 
	\begin{equation}
		\begin{aligned}
			\mathcal{L}_Y &\supset-[V_{ud}Y^{D, dd}_2\bar u P_R d - Y^{U,uu*}_2 V_{ud}\bar u P_L d] H^+ + H.C\\
			& \supset-\frac{1}{2}\left[\bar u (a_{ud}+b_{ud}\gamma_5)d\right]H^+ + H.c
		\end{aligned}
	\end{equation}
	where 
	\begin{equation}
		\begin{aligned}
			&a_{ud}=V_{ud}(Y_{2}^{D, dd}-Y_{2}^{U,uu*}),\ \ \ \ b_{ud}=V_{ud}(Y_{2}^{D, dd} +Y_{2}^{U,uu*}).
		\end{aligned}
	\end{equation}
	
	The (u, d) quark EDM and CDM from charged Higgs are expressed as 
	\begin{equation}
		\begin{aligned}
			d^{\mathrm{c}}_{uEDM}&=\frac{|V_{ud}|^2|Y_{2}^{D, dd}||Y_{2}^{U, uu}|m_d e}{(4\pi)^2m_{H^+}^2}\mathrm{sin}(\theta^{uu}+\theta^{dd})\left[Q_dg(y_d)+f(y_d)\right],\\
			d_{dEDM}^{\mathrm{c}}& =\frac{|V_{ud}|^2|Y_{2}^{D, dd}||Y_{2}^{U, uu}|m_u e}{(4\pi)^2m_{H^+}^2}\mathrm{sin}(\theta^{uu}+\theta^{dd}) [Q_ug(y_u)-f(y_u)],\\
			d_{uCDM}^{\mathrm{c}}& = \frac{|V_{ud}|^2|Y_{2}^{D, dd}||Y_{2}^{U, uu}|m_d }{(4\pi)^2m_{H^+}^2}\mathrm{sin}(\theta^{uu}+\theta^{dd})g(y_d), \\
			d_{dCDM}^{\mathrm{c}}& = \frac{|V_{ud}|^2|Y_{2}^{D, dd}||Y_{2}^{U, uu}|m_u }{(4\pi)^2m_{H^+}^2}\mathrm{sin}(\theta^{uu}+\theta^{dd})g(y_u),
		\end{aligned}
	\end{equation}
	
	where $y_q = m^2_q/m^2_{H^\pm}$.
	\begin{eqnarray}
		g(x) = -{1\over 2(1-x)^2}(3-x+ {2lnx \over 1-x})\;\;\;\;f(x) = {1\over 2(1-x)^2} (1+x + {2xln x\over 1-x})
	\end{eqnarray}
	
	The neutral EDM due to the charged Higgs is 
	\begin{equation}
		\begin{aligned}
			d_n^{\mathrm{c}}& = \eta_{EDM}d^{\mathrm{c}}_{nEDM}(\mu_h)+\left(\frac{11}{3}\eta_{CDM}-\frac{8}{3}\eta_{EDM}\right)d^{\mathrm{c}}_{nCDM}(\mu_h).\\
		\end{aligned}
	\end{equation}
	Since the masses of the u and d quarks are much smaller than that of the charged Higgs ,$y_q\ll1$ ,  we have $f(y_q)\approx1/2$ . For $m_{H^{\pm}}\in(0.3 , 0.7)\ \mathrm{TeV}$,  $m_u=2.16\ \mathrm{MeV}$ and $m_d =4.70\ \mathrm{MeV}$~\cite{ParticleDataGroup:2024cfk}, the range of $g(y_d)$ is about $(20.6 , 22,3)$ and the range of $g(y_u)$ is about $(22.2,23.9)$. Therefore, $f(y_q)$ and $g(y_q)$ vary slightly with $m_{H^{\pm}}$.  
	
	In the following we present our numerical results. To make contact with the purpose of the dark matter model under study, we use $Y_2^{uu, dd}$ in Eq.~\eqref{yukawa} and replace $Y^{qq}_2$ by  $Y^{qq}_2\cos\theta^{qq}$ as mentioned before. The results are shown in Fig.~\ref{quarkEDM neutral mA} and Fig.~\ref{quarkEDM lambda3}.  
	
	In Fig.~\ref{quarkEDM neutral mA}, we show the u and d quark contribution to the neutron EDM due to the neutral Higgs exchange divided by $\mathrm{tan}\theta^{qq}$ as a function of $m_H/m_A$ at different $|Y^{D,dd}_2|$. The left and right figures represent $m_H = 0.5\ \mathrm{TeV}$ and $1\ \mathrm{TeV}$ respectively. When $m_H = m_A$, the contributions from $H$ and $A$ to the $u$- and $d$-quark EDMs cancel each other out, leading to a vanishing neutron EDM. As long as $m_A$ and $m_H$ are sufficiently close, corresponding $\Delta m_s^{NP}$ small enough, these contribution can satisfy current experimental limits.  But when $m_H/m_A$ departs from 1, the contribution to EDM can be substantial. In the left panel of the Fig. \ref{quarkEDM lambda3}, we show the $|d^{\mathrm{n}}_{qEDM}|/\mathrm{tan}\theta^{qq}$ as a function of $|Y^{D,dd}_2|$ with $m_H/m_A = 1.05$. We can find that the contribution of the d-quark is approximately one order of magnitude greater than that of the u-quark when $m_H$ is the same. The neutron EDM contribution from neutral Higgs exchange scales as $1/m_H^2$, therefore a lighter $H$ boson leads to a larger EDM which has shown in the Fig.~\ref{quarkEDM neutral mA} and Fig~\ref{quarkEDM lambda3}. Besides, one can easily produce a neutron EDM as large as the experimental limits. Taking $m_H/m_A = 1.05$ , $m_H=0.5\ \mathrm{TeV}$ and  $Y^{D,dd}_2 = 1$ for example, we find that to satisfy the experimental bound $|d_n| < 1.8\times 10^{-26}\ e\cdot\mathrm{cm}$ , the $\tan\theta^{qq}$ would need to be less than about $10^{-1}$. We can also find that the contributions of u quark and d quark have opposite signs.  Here we used $Y^{D,dd}_2 = 1$ for discussion. If the coupling can reach the perturbative bounds of $\sqrt{4 \pi}$ or $\sqrt{8\pi/3}$, the limit will be proportionally stronger. 
	
	Regarding the charged Higgs contribution, the cancellation between the quark EDM and CEDM does not occur, leading to potentially larger effects for the same set of parameters. Furthermore, as there is only a single charged Higgs boson involved, the $H$-$A$ cancellation characteristic of the neutral sector is also absent.  In the right panel of Fig.~\ref{quarkEDM lambda3}, using $m_{H^+} = 0.5\ \mathrm{TeV}$ and $m_{H^+} = 1\ \mathrm{TeV}$ we find that the charged Higgs boson contribution is about two orders of magnitude larger than that from neutral Higgs boson. Taking $Y^{D,dd}_2 = 1$ and $m_{H^+}= 0.5\ \mathrm{TeV}$ , it is found that $|\tan\theta^{uu} + \tan\theta^{dd}|$ should be smaller than $10^{-3}$ to satisfy the experimental constraint. Future improvements in experimental sensitivity will further constrain this parameter space.
	
	As shown in Fig. \ref{quarkEDM neutral mA}, the d-quark and u-quark contributions are opposite in sign.  Therefore, when $\theta^{uu}$ and $\theta^{dd}$ have opposite signs, the contributions from the charged and the neutral scalar may also be opposite in sign. This cancellation allows the total neutron EDM to satisfy experimental constraint even for large values of $\theta^{uu}$ and $\theta^{dd}$. Taking $m_H/m_A = 1.05$, $m_H=m_{H^+}=0.5\ \mathrm{TeV}$ and $|Y^{D,dd}_2|=1$ as our benchmark example, we have $d_{n}^{\mathrm{n}}\sim10^{-25}(\mathrm{tan}\theta^{uu}-\mathrm{tan}\theta^{dd})$ and $d_{n}^{\mathrm{c}}\sim10^{-23}(\mathrm{tan\theta^{uu}+\mathrm{tan}\theta^{dd}})$. The cancellation occurs when $|\mathrm{tan}\theta^{uu}-\mathrm{tan}\theta^{dd}|$ is approximately two orders of magnitude larger than $|\mathrm{tan}\theta^{uu}+\mathrm{tan}\theta^{dd}|$.

	\section{Conclusions\label{concl}}

	Based on the type-III two Higgs doublet model, we propose a renormalizable framework incorporating light dark matter to accommodate the recent possible excesses in  $B^+\to K^+\nu\bar\nu$ and $K^+\to\pi^+\nu \bar \nu$ reported by Belle II and NA62, respectively. We show that this renormalizable model will realize the effective Lagrangian in the previous paper. Besides, we study the implications of this model for the $B_s - \bar B_s$ mixing and the neutron EDM. Within the allowed parameter space, we find non-negligible contributions to $B_s - \bar B_s$ mixing. The NP contribution can reach a maximum size as low as about $-0.13\ (-0.08)\ \mathrm{ps}^{-1}$ for $|Y^{D,dd}_2|=\sqrt{4\pi}\ (\sqrt{8\pi/3})$, which is significant compared to $\Delta M_s^{NP}=-0.465\pm 0.63\ \mathrm{ps}^{-1}$. If more precise SM and experimental values in the future narrow down the allowed region, constraints may be imposed on the maximal allowed $Y_2^{D, dd}$. We also study the neutron EDM induced by both neutral and charged Higgs exchange, by allowing for CP-violating phases in the first-generation Yukawa couplings, $Y^{D, dd}_2$ and $Y^{U, uu}_2$. We find that there is a cancellation due to the exchange of neutral spin-zero particle, but QCD renormalization group evolution will lift this cancellation which in fact is generally true for any neutral Higgs contribution. In this model, neutron EDM will also receive contributions from charged Higgs boson $H^\pm$. Although the charged Higgs boson contribution is nominally two orders of magnitude larger than neutral Higgs boson contribution, they can become comparable in magnitude and opposite because of the phase, leading to the cancellation. The allowed CP-violating phases in the Yukawa sector can generate a neutron EDM at a level consistent with current bounds.

	\section*{Acknowledgments}
	
	This work was supported by the Fundamental Research Funds for the Central Universities, by the National Natural Science Foundation of the People’s Republic of China (Nos. 12090064, 12375088, 123B2079 and W2441004 ).

	\setlength{\bibsep}{.15\baselineskip plus 0.1ex}
	
	\bibliographystyle{unsrt} 
	\bibliography{refs.bib}

\end{document}